\documentclass[twocolumn,
aps,nofootinbib,showpacs,showkeys,preprint
tightenlines,preprintnumbers]{revtex4}

\usepackage{epsf,epsfig,subfigure,graphicx,amsmath,amssymb}
\usepackage{amsfonts}
\usepackage{latexsym}
\usepackage{color}

\newcommand{\vew}{v_{\rm EW}} 

\newcommand{\dis}[1]{\begin{equation}\begin{split}#1\end{split}\end{equation}}
\newcommand{\eq}[1]{Eq.~(\ref{#1})}

\newcommand{\bfrac}[2]{{\left(\frac{#1}{#2} \right)  }}

\newcommand{\VEV}[1]{\langle #1 \rangle}
\newcommand{\OVER}[1]{\,\overline{\hskip -0.5mm #1}}

\newcommand\lsim{\mathrel{\rlap{\lower4pt\hbox{\hskip1pt$\sim$}}
    \raise1pt\hbox{$<$}}}
\newcommand\gsim{\mathrel{\rlap{\lower4pt\hbox{\hskip1pt$\sim$}}
    \raise1pt\hbox{$>$}}}

\newcommand\etal{{\it et al.}}

\newcommand\ie{{\it i.e.}~}
              
\newcommand\mnlsp{m_{\rm NLSP}}

\newcommand\squark{\widetilde q}

\newcommand{\SUC}{{SU(3)$_c$}}
\newcommand{\SUW}{{SU(2)$_L$}}
\newcommand{\UY}{{U(1)$_Y$}}

\newcommand\gluino{\tilde G}
\newcommand\mgluino{m_{\gluino}}

\newcommand\treh{T_{\rm R}}     
\newcommand\Tdec{T_{\rm dec}}   

\newcommand\fa{f_{a}}

\newcommand\tev{\,{\rm TeV}}
\newcommand\gev{\,{\rm GeV}}
\newcommand\mev{\,{\rm MeV}}
\newcommand\kev{\,{\rm keV}}

\newcommand\axino{{\tilde{a}}}
\newcommand\maxino{{m_{\axino}}}

\newcommand\omegaantp{\Omega^{\rm NTP}_{\axino}}

\newcommand\mgravitino{{m_{3/2}}}

\begin{document}
\draft

\title{
Review of axino dark matter\\
}

\author{Ki-Young Choi,$^{(1,2)}$ Jihn E.  Kim,$^{(3)}$ Leszek Roszkowski\,$^{(4)}$\footnote{On leave of absence from the University of Sheffield, UK.}}
\affiliation
{
$^{(1)}$Asia Pacific Center for Theoretical Physics, Pohang, Gyeongbuk 790-784, Republic of Korea,  \\
$^{(2)}$Department of Physics, POSTECH, Pohang, Gyeongbuk 790-784, Republic of Korea\\
$^{(3)}$Department of Physics, Kyung Hee University,\\
Seoul 130-701, Korea,\\
$^{(4)}$National Centre for Nuclear Research, Ho\.za 69, 00-681, Warsaw, Poland
}

%

\begin{abstract}
  We review the status of axino dark matter. Two hierarchy problems,
  the strong CP problem and the gauge hierarchy problem, have led to
  introducing into particle physics a spontaneously broken global
  Peccei-Quinn symmetry and a softly broken supersymmetry,
  respectively. Combining them implies the presence of not only
  an axion, but also of its scalar component, saxion, and their
  fermionic partner, axino. Among these, the axion and the axino are
  attractive dark matter candidates. Various possibilities for the
  axino as dark matter are discussed.

\end{abstract}


\preprint{}

\vspace*{3cm}
\maketitle


\section{Introduction}
\label{sec: introduction}
The axino, stable or almost stable on cosmological time scales, is
a well-motivated dark matter (DM) candidate. Axinos differ from ordinary
weakly interacting massive particles (WIMPs) in that they are
exceedingly weakly interacting (EWIMPs), which radically changes their
cosmological properties and also ways of testing them in experiment.

Relic axinos can be produced in a hot plasma or in decays of heavy
particles in the early universe. The most interesting case is when it
can be cold dark matter (CDM) - the possibility that was first
considered in Ref.~\cite{CKR00} where axinos were generated in the
decays of heavier particle after freezeout in the process that was
dubbed {\it non-thermal production} (NTP). This was next extended in
Ref.~\cite{CKKR01} to include thermal production (TP) through
scatterings and decays of particles in thermal equilibrium. It was
also pointed out there that the axino could as well be hot DM (HDM) or
warm DM (WDM), or that, for instance, one of its populations could be
warm while the other cold. Since then the axino as DM in
cosmology, astrophysics and collider experiments has been studied in
many papers~\cite{CRS02,CRRS04,Steffen04,Strumia10,AxinoRevs,flaxino,
  Baer,Baer:2010wm,Baer_mixed, Wyler09,Baer:2011uz,Kang:2008jq}.

The axino as the superpartner of the axion was first considered right after it was
recognized that softly broken supersymmetry (SUSY) was relevant for
particle physics~\cite{NillesRaby82,Tamv82,Frere83}.  Therefore, an
experimental confirmation of axino existence would validate two
theoretical hypotheses designed to solve two respective hierarchy
problems: the strong CP problem by a very light
axion~\cite{KimRMP10} and the gauge hierarchy problem by
SUSY~\cite{Nilles84}.

Similarly to the axion, axino interactions with Standard Model (SM)
and minimal supersymmetric SM (MSSM) particles are suppressed by the
axion decay constant $\fa$. On the other hand, the quantity that is
most relevant for the axino in astro-particle physics, and at the same
time most poorly known, is its mass $\maxino$.  Several theoretical
calculations of the axino mass
followed~\cite{ChunKN92,Chun:1995hc,Kim:2012bb}.  A method for
calculating axino mass applies to any goldstino (the superpartner of a
Goldstone boson). A goldstino related to the Goldstone boson has a
root in a global U(1) symmetry and receives its mass below the SUSY
breaking scale. SUSY breaking triggers the super-Higgs mechanism and
is related to the gravitino mass $\mgravitino$ which has recently been clarified
in~\cite{Kim:2012bb}. Even though generically axino mass is of order
$\mgravitino$, a theoretically allowed mass range encompasses all the
range relevant for hot, warm and cold DM axinos. We will discuss this
in more detail below.

In the early days, a very light HDM-like axino from the decay of a
photino was considered~\cite{Masiero84} to constrain the photino mass
dependence on the axion decay constant $\fa$. Rajagopal, Turner and
Wilczek~\cite{RTW91} considered axinos in a keV mass range. Axinos in
this mass range can give a right amount of DM if produced in
the thermal equilibrium and constitute WDM in the standard Big Bang
cosmology. However, this kind of thermal axino is not useful if the
reheating temperature $\treh$ after inflation is much lower than the
Peccei-Quinn (PQ) symmetry breaking scale $\fa$. In this case the
population of primordial axinos is strongly diluted by cosmic
inflation  and axinos are subsequently re-generated after reheating.

Due to the exceedingly small interaction strength, $1/\fa$, the relic
abundance of thermal axinos depends on the reheating temperature and
on the axion model. This special feature allows us to have a glimpse
on the earliest time after inflation, through the reheating
temperature inferred from the relic density of axino DM. It is also
worth mentioning that, due to the strongly suppressed interaction
strength, it is not necessary to assume $R$ parity for axinos to
constitute DM.

In this paper, we provide a review of the axino as DM and its
implication in cosmology, astro-particle physics and collider
phenomenology.\footnote{For other recent reviews on axinos as DM
  particles, see~\cite{AxinoRevs}.}

In Sect.~\ref{sec:DefA}, we introduce axino frameworks and present an
effective Lagrangian describing its interactions. In
Sect.~\ref{sec:AxinoCosmology}, we discuss in more detail the
production of axinos, as well as cosmological applications for the
HDM, WDM, CDM, and superheavy axino cases. In
Sect.~\ref{sec:AxinoColliders}, we review CDM axino production at
colliders and its links with the early Universe. In
Sec.~\ref{sec:conclusions}, we summarize the discussion of axino DM
and its cosmological consequences.

\section{Axion/axino models}
\label{sec:DefA}

The strong $CP$ problem can most naturally be solved by introducing a very light
axion field $a$~\cite{KimRMP10} which couples to the gluon anomaly
\dis{
\mathcal{L} = \frac{\alpha_s a}{8\pi \fa}\,G^a_{\mu\nu}\widetilde{G}^{a\,\mu\nu},
\label{aGG}
}
where $\alpha_s=g_s^2/4\pi$ is the strong coupling constant and
$\widetilde{G}^{a\,\mu\nu} = \frac12 \epsilon^{\mu\nu\rho\sigma }
G^a_{\rho\sigma}$ is the dual of the field strength $G^{a\,\mu\nu}$
for eight gluons $G^a_\mu\,(a=1,2,\cdots,8)$. This interaction term
can be obtained after integrating out colored heavy fields below the
PQ symmetry breaking scale $\fa$ but above the electroweak scale
$\vew$. The axion decay constant $\fa$ is constrained from
astrophysical and cosmological considerations to a narrow window
$10^{10}\gev \lsim \fa \lsim 10^{12}\gev$~\cite{KimRMP10}. The upper
bound is obtained here assuming that the initial misalignment angle is
of order one, and can be lifted if the angle were assumed smaller than
one~\cite{BaeHuhKim09}.

A general low energy axion interaction Lagrangian can be written in
terms of the effective couplings with the SM fields $c_1$, $c_2,$ and
$c_3$, which arise after integrating out all heavy PQ-charge carrying
fields.  The resulting effective axion interaction Lagrangian terms
are~\cite{KimRMP10}
\dis{
\mathcal{L}&^{\rm eff}_{\rm int} \,=\,c_1 \frac{(\partial_\mu a)}{\fa}
\sum_q \bar{q}  \gamma^\mu\gamma_5q \\
& - \sum_q  (\bar{q}_L m q_R e^{ic_2 a/\fa }+ \textrm{h.c.} )
+\frac{c_3}{32\pi^2 \fa} a G\widetilde{G}\\
 & + \frac{C_{aWW}}{32\pi^2 \fa}
 a W\widetilde{W} +\frac{C_{aYY}}{32\pi^2 \fa}
 a Y\widetilde{Y}  + \mathcal{L}_{\rm leptons},
 \label{eq:efflagr}
}
where $c_3$ can be set to one by rescaling $\fa$. The axion decay
constant $f_S$, $\theta= a/f_S$, is defined up to the domain wall
number, $f_S=N_{\rm DW } \fa$. The derivative interaction term
proportional to $c_1 $ preserves the PQ symmetry. The $c_2$-term is
related to the phase of the quark mass matrix, and the $c_3$-term
represents the anomalous coupling.  The axion-lepton interaction term
$\mathcal{L}_{\rm leptons}$ is analogous to the axion-quark
interaction term.

Two prototype field theory models for {\it very light} axions have
been considered in the literature.  At the SM level, one considers the
six SM quarks, $u, d, s, \ldots$, as strongly interacting matter
fermions. Above the electroweak scale $\vew\simeq 247\,\gev$ one
additionally introduces beyond the SM (BSM) heavy vector-like quarks
($Q_i, \overline{Q}_i$), which in the interaction
Lagrangian~(\ref{eq:efflagr}) are next integrated out.

At the field theory level the axion is present if there exist quarks
carrying the net PQ charge $\Gamma$ of the global U(1)$_{\rm PQ}$
symmetry.  In the Kim-Shifman-Vainstein-Zakharov~(KSVZ)
model~\cite{KSVZ79} one introduces only heavy quarks as PQ charge
carrying quarks. This results in $c_1=c_2=0$, and $c_3=1$ below the
$\vew$, or below the QCD scale $\Lambda_{\rm QCD}$.
The gluon anomaly term (the $c_3$ term), induced by an effective heavy
quark loop, then solves the strong $CP$ problem.  The axion field is a
component of the SM singlet scalar field $S$.  In the
Dine-Fischler-Srednicki-Zhitnitskii (DFSZ) model~\cite{DFSZ81} instead one does not
assume any net PQ charge in the BSM sector, and instead the SM quarks
are assigned the net PQ charge, \ie, $c_1=c_3=0$ and
$c_2\ne 0$ below the electroweak scale $\vew$. Here also, the axion is
predominantly a part of the SM singlet scalar field $S$. Several
specific implementations of the KSVZ and DFSZ frameworks can be found
in Refs.~\cite{Kim98,KimRMP10} which, however, require a whole host of
additional BSM fields. Therefore, any
references to the properties of the KSVZ and the DFSZ models can serve
at best as just guidelines.  In this respect, it is unfortunate that
there exists only one reference clarifying the axion-photon-photon
coupling from a string-derived BSM framework~\cite{IWKim06}.

SUSY models of very light axion can provide a clue about the magnitude
on the axion decay constant $\fa$. In Ref.~\cite{Kim83} it was
speculated that it is related to the MSSM Higgs/higgsino mass
parameter $\mu$ as $f_a\sim \sqrt{\mu M_P}$. Recently, a permutation
symmetry $S_2\times S_2$ has been used to relate $\fa$ and $\mu$,
realizing the old hypothesis in terms of a discrete
symmetry~\cite{Kim13}. Using a discrete symmetry toward an approximate
global symmetry is theoretically welcome in that it evades the
wormhole breaking of the PQ symmetry~\cite{Barr92}.

Before one considers spontaneous symmetry breaking of U(1)$_{\rm PQ}$,
the axion Lagrangian can be said to have the axion shift symmetry
(which is just a phase rotation), $a\rightarrow a + \textrm{const}$,
and the physical observables are invariant under the PQ phase
rotation. Below $\fa$, the PQ rotational symmetry is broken, which is
explicitly reflected as a breaking of the axion shift symmetry through
the appearance of the $c_2$ and $c_3$ terms in
Eq.~(\ref{eq:efflagr}). However, the $c_2$ term enters into the phase and a
discrete shift of the axion field can bring it back to the original
value. The $c_3$ term is the QCD vacuum angle term and if the vacuum
angle is shifted by $2\pi$ then it comes back to the original
value. Thus, even though the U(1)$_{\rm PQ}$ is broken, one of its
discrete subgroups, \ie, the one corresponding to the common
intersection of the subgroups corresponding to the $c_2$ and $c_3$
terms, can never be broken. As a result, the combination $c_2+c_3$ is
invariant under the axion shift symmetry, and $c_2+c_3$ is defined to
be an integer signifying the unbroken discrete subgroup of U(1)$_{\rm
  PQ}$~\cite{KimRMP10}. It is called the domain wall number $N_{\rm
  DW}=|c_2+c_3|$~\cite{Sikivie:1982qv}.

When an axion model is
supersymmetrized~\cite{Tamv82,NillesRaby82,Frere83}, there appears a
fermionic SUSY partner of the axion field called the {\it axino} $\axino$,
as well as a real scalar field $s$ named the {\it saxion}. Together with
the axion they form an axion
supermultiplet $A$
\dis{ A=\frac{1}{\sqrt2}(s+ia)+\sqrt2 \axino
  \vartheta + F_A \vartheta\vartheta,
}
where $F_A$ stands for an auxiliary field of $A$
and $\vartheta$ for a Grassmann coordinate.  The
interaction of the axion supermultiplet is obtained by
supersymmetrizing the axion interaction in \eq{eq:efflagr}.  In
particular, the interaction of the axion supermultiplet $A$
with the vector multiplet $V_a$, which is a SUSY version of the $c_3$ term in
\eq{eq:efflagr}, is given by
\dis{ {\mathcal L}^{\rm eff}
  =-\sum_{V}\frac{\alpha_V\, C_{aVV}}{2\sqrt2\pi \fa}\int A\,{\rm
    Tr}\,[V_a V^a] + \textrm{h.c.},
\label{Leff3}
}
where $\alpha_V$ denotes a gauge coupling, $C_{aVV}$ is a model dependent
constant and the sum goes over the SM gauge groups.
From this the relevant axino-gaugino-gauge boson and axino-gaugino-sfermion-sfermion
interaction terms can be derived and are given by~\cite{Choi:2011yf}
\dis{
{\mathcal L}^{\rm eff}_\axino&=i\frac{\alpha_s}{16\pi \fa}\overline{\axino}
\gamma_5[\gamma^\mu,\gamma^\nu]\tilde{G}^b
G^b_{\mu\nu} +\frac{\alpha_s}{4\pi \fa}\overline{\axino}\tilde{g}^a
\sum_{\tilde{q}}g_s \tilde{q}^*T^a\tilde{q}\\
&+i\frac{\alpha_2C_{aWW}}{16\pi \fa}\overline{\axino}\gamma_5[\gamma^\mu,\gamma^\nu]\tilde{W}^b
W^b_{\mu\nu}\\
&+\frac{\alpha_2}{4\pi \fa}\overline{\axino}\tilde{W}^a
\sum_{\tilde{f}_D}g_2 \tilde{f}_D^*T^a\tilde{f}_D\\
 &+ i\frac{\alpha_YC_{aYY}}{16\pi \fa}\overline{\axino}
 \gamma_5[\gamma^\mu,\gamma^\nu] \tilde{Y}Y_{\mu\nu}\\
&+\frac{\alpha_Y}{4\pi \fa}\overline{\axino}\tilde{Y}
\sum_{\tilde{f}}g_Y \tilde{f}^*Q_Y\tilde{f},
\label{eq:Laxino}
}
where the terms proportional to $\alpha_2$ correspond to the \SUW~ and
the ones proportional to $\alpha_Y$ to the \UY~ gauge groups,
respectively.  $C_{aWW}$ and $C_{aYY}$ are model-dependent couplings
for the \SUW~ and \UY~ gauge group axino-gaugino-gauge boson anomaly
interactions, respectively, which are defined after the standard
normalization of $\fa$, as in  \eq{aGG} for the \SUC\ term. Here
$\alpha_2$, $\widetilde{W}$, $W_{\mu\nu}$ and $\alpha_Y$,
$\widetilde{Y}$, $Y_{\mu\nu}$ are, respectively, the gauge coupling,
the gaugino field and the field strength of the \SUW~ and \UY~ gauge
groups. $\tilde{f}_D$ represents the sfermions of the \SUW-doublet and
$\tilde{f}$ denote the sfermions carrying the \UY\ charge.

Similarly, one can derive supersymmetrized interactions of the axion
supermultiplet with matter multiplet as a generalization of the $c_1$
and $c_2$ terms in \eq{eq:efflagr}. Ref.~\cite{Bae11} considered a
generic form of effective interactions and clarified the issue of
energy scale dependence of axino interactions.  At some energy scale
$p$ which is larger than the mass of the PQ-charged and gauge-charged
multiplet $M_\Phi$, the axino-gaugino-gauge boson interaction is
suppressed by $M_\Phi^2/p^2$. This suppression is manifest in the DFSZ
axion model or in the KSVZ model if the heavy quark mass is relatively
low compared to the PQ scale in which case the heavy quark is of
course not integrated out.

However, SUSY must be broken at low energy and thus a SUSY relation
between the axino and the axion is modified.  In fact, the most
important axino parameter in cosmological considerations, the axino
mass $\maxino$, does not even appear in Eq.~(\ref{eq:Laxino}). SUSY
breaking generates the masses for the axino and the saxion and
modifies their definitions. The saxion mass is set by the SUSY soft
breaking mass scale, $M_{\rm SUSY}$~\cite{Tamv82,Nieves:1985fq}. The
axino mass, on the other hand, is strongly model dependent.  An
explicit axino mass model with SUSY breaking was first constructed long
time ago~\cite{Kim83} with the superpotential for the PQ symmetry
transformation $S\to e^{i\alpha} S$ and $\OVER{S}\to
e^{-i\alpha}\,\OVER{S}$,
\dis{ W=\sum_{i=1}^{n_I} Z_i(S
  \OVER{S}-f_i^2),~n_I\ge 2.\label{eq:SUSYglobBr}
}
With $n_I=1$, the
U(1)$_{\rm PQ}$ symmetry is spontaneously broken but SUSY remains
unbroken. The case $n_I=2$ was also considered in~\cite{Kim83} but this model
gives $\maxino=0$.

As first pointed out by Tamvakis and Wyler~\cite{Tamv82}, the axino
mass is expected to receive at least a contribution at the order of
$\maxino \sim \mathcal{O}(M_{\rm SUSY}^2/\fa)$ at tree level in the
spontaneously broken global SUSY. In the literature, a whole range of
the axino mass was considered; and in fact it can be even much
smaller~\cite{Frere83,Masiero84,Moxhay:1984am,ChunKN92,Goto:1991gq},
or much larger, than the $M_{\rm SUSY}$~\cite{Chun:1995hc}.  Because
of this strong model dependence, in cosmological studies one usually
assumes the axino interactions from the U(1)$_{\rm PQ}$ symmetry and
treats axino mass as a free parameter.

Recently, the issue of a proper definition of the axion and the axino
was studied in the most general framework, including non-minimal
K$\ddot{\rm a}$hler potential~\cite{Kim:2012bb}. In that study, axino
mass is given by $\maxino=\mgravitino$ for $G_A=0$,
where $G=K+\ln |W|^2$ and $G_A\equiv \partial G / \partial A$. For $G_A\neq 0$,
axino mass depends on the
details of the K$\ddot{\rm a}$hler potential, and it was shown that
the case given by Eq.~(\ref{eq:SUSYglobBr}) belongs to one of these
examples. In gauge mediation scenario, the gaugino mass is the
dominant axino mass parameter. In the case of gravity mediation, the axino mass
is likely to be greater than the gravitino mass but one cannot rule out
lighter axinos~\cite{Kim:2012bb}.

\begin{figure}[t]
  \begin{center}
  \begin{tabular}{c}
   \includegraphics[width=0.45\textwidth]{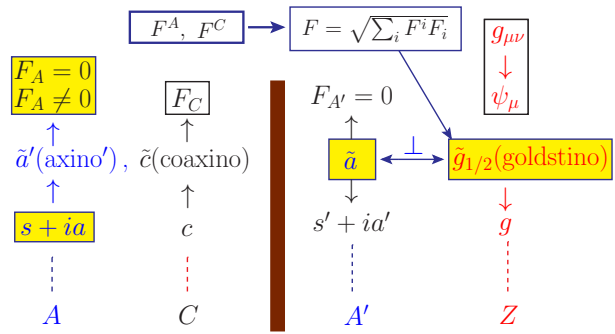}
  \end{tabular}
  \end{center}
  \caption{ Axion (blue) and goldstino (red) multiplets.
The axion direction $a$ is defined by the PQ
    symmetry and the goldstino ($\tilde{g}_{1/2}$) and axino ($\tilde a$)
    directions are defined by the fermion mass eigenvalues.  The
    primed fields are not mass eigenstates.}
\label{fig:AxinoDef}
\end{figure}

One lucid but often overlooked aspect of the axino is that its
definition must be given at a mass eigenstate level.  The coupling to
the QCD sector given in the first line of Eq.~(\ref{eq:Laxino}) can
plausibly be that of the axino but it does not give the axino
mass. This is because the axino is connected to two kinds of symmetry
breaking, the PQ global symmetry breaking and the SUSY breaking, which
in general are not orthogonal to each other. The PQ symmetry breaking
produces an almost massless pseudo-Goldstone boson (the axion), while
SUSY breaking produces a massless goldstino. The massless goldstino is
then absorbed into the gravitino to make it heavy via the super-Higgs
mechanism. This raises the question of what the axino really is. This
issue is shown in Fig.~\ref{fig:AxinoDef} taken from
Ref.~\cite{Kim:2012bb}. The axino must be orthogonal to the massless
goldstino component. Therefore, for the axino to be present in a
spontaneously broken supergravity theory, one has to introduce at
least two chiral fields~\cite{Kim83}. Even though its name refers to
the axion-related QCD anomaly, one must select the component which is
orthogonal to the goldstino. If there are two SM singlet chiral
fields, this is easy since there is only one component left beyond the
goldstino. However, if more than two chiral fields are involved in
SUSY breaking, more care is needed to identify the orthogonal mass
eigenstate.  Among the remaining mass eigenstates beyond the
goldstino, a plausible choice for the axino field is the component
whose coupling to the QCD anomaly term is the biggest. For two initial
chiral fields in Fig.~\ref{fig:AxinoDef}, $\axino'$ has the anomaly
coupling of Eq.~(\ref{eq:Laxino}) and hence the $\axino$ coupling to the
QCD sector is equal to or smaller than those given in
Eq.~(\ref{eq:Laxino}). The remaining coupling is the one to the $s=\pm\frac12$
components of a massive gravitino. Therefore, for the two initial chiral
fields,  axino cosmology must include the gravitino, as well, if
$\axino'$ is not identical to $\axino$. The ``leakage'' is parametrized by
the $F$-term of the initial axion multiplet $A$. With more than two
initial chiral fields, the situation involves more mass parameters.
One notable corollary of Ref.~\cite{Kim:2012bb} is that the axino CDM
relic abundance for $\maxino<\mgravitino$ is an over-estimation if $A$ obtains
the $F$-term.

The saxion mass is most likely of order $\mgravitino$, and therefore
saxions decay to SM particles. A notable cosmological implications
of the saxion decay is known to be entropy production and a dilution of cosmic particles and
the cosmic energy density. When applied to axion cosmology, the effect
leads to some increase of the cosmological upper bound on
$\fa$~\cite{Kim91, KimHB96,Kawasaki:2007mk,Kawasaki:2011ym}.  Axions
and axinos produced from the decays of saxions can also affect the cosmic
microwave background temperature anisotropy by contributing an additional
relativistic component~\cite{Hasenkamp:2011em,Choi:2012zna,Graf:2012hb,Bae:2013qr,Graf:2013xpe}.

\begin{figure}[t]
  \begin{center}
  \begin{tabular}{c}
   \includegraphics[width=0.45\textwidth]{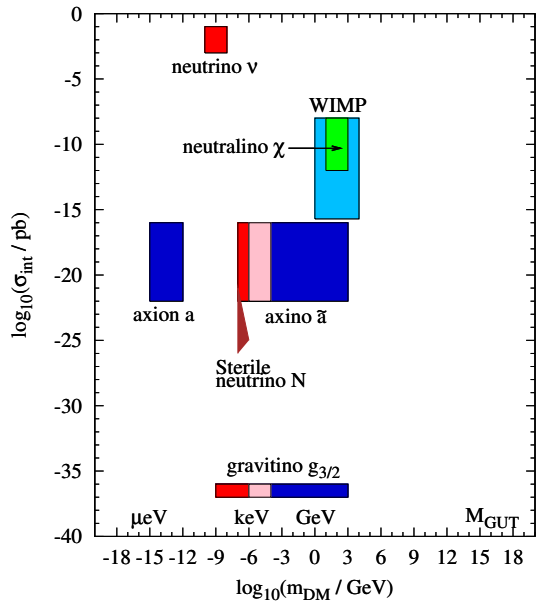}
  \end{tabular}
  \end{center}
  \caption{Several well-motivated candidates of DM are
    shown. $\sigma_{\rm int}$ is the typical strength of the
    interaction with ordinary matter. The red, pink and blue colors
    represent HDM, WDM and CDM, respectively. We updated the
      previous figures~\cite{CMSSM-dm,KimRMP10} by including the
      sterile neutrino
      DM~\cite{Boyarsky:2005us,Boyarsky:2009ix,Abazajian:2012ys}.  }
\label{fig:DMtype}
\end{figure}
\section{Axinos in cosmology}\label{sec:AxinoCosmology}

In this section we proceed to review axinos as relics in
cosmology. We will treat axino mass as a free parameter ranging from
eV to multi-TeV scales. A schematic representation of the strength and
mass of axino DM is shown in Fig.~\ref{fig:DMtype}, which is an
updated version of the figure originally included in
Ref.~\cite{CMSSM-dm} and next modified in~\cite{KimRMP10}. Marked schematically
in the figure are also other well motivated candidates for DM, with
HDM, WDM and CDM classes of candidates shown in red, pink and blue
colors, respectively. As one can see, depending on axino mass and
production mechanism, cosmic axinos may as well fall into one or
actually more than one (e.g., CDM and WDM) population of relics, as
discussed below.

\subsection{Production of relic axinos}\label{subsec:CosProd}

As stated in Introduction, there are two generic ways of producing
relic axinos in the early Universe: thermal production from
scatterings and decays of particles in thermal equilibrium, and
non-thermal production from the decays of heavier particles after their
freezeout.

\vskip 0.5cm
\subsubsection{Thermal production}

Primordial axinos decouple from thermal equilibrium at the temperature~\cite{RTW91}
\dis{
\Tdec= 10^{11}\,\gev \bfrac{\fa}{10^{12}\gev}^2\bfrac{0.1}{\alpha_s}^3.\label{eq:DecTemp}
}
They overclose the Universe unless their mass is bounded to be smaller
than $\kev$~\cite{RTW91}.  In inflationary cosmology, the population
of primordial axinos is strongly diluted by cosmic inflation; however
axinos are re-generated during reheating.  When the reheating
temperature $\treh$ is below the decoupling temperature, axinos do not
reach the equilibrium level. However, axinos can be produced from the
scatterings in thermal plasma, and the number density is proportional
to $\treh$, in which case the $~\kev$ mass upper bound of
Ref.~\cite{RTW91} is relaxed. The calculation follows the same line of
logic which was used for the gravitino regeneration and
decay~\cite{Ellis:1984eq,Kawasaki:1994af}.  If the axino mass is
between around an MeV to several GeV, the correct axino CDM density is
obtained with $\treh$ less than about $5\times
10^4\gev$~\cite{CKKR01}.

Thermal production of axinos is described by the Boltzmann equation
where the first term on the r.h.s. corresponds to scatterings and the
second one to decays~\cite{CKKR01,Steffen04,Strumia10,Gomez:2008js},
\dis{
  \frac{d n_\axino}{d t} + 3 H n_\axino &= \sum_{i,j} \VEV{\sigma (i+j
    \rightarrow \axino+\ldots) v_{\rm rel}} n_i n_j\\
  &+ \sum_{i} \VEV{\Gamma (i\rightarrow \axino +\ldots) } n_i,
\label{Boltzmann}
}
where $H$ denotes the Hubble parameter, $\sigma (i+j \rightarrow
\axino+\ldots) $ is the scattering cross section for particles $i,j$
into final states involving axinos and $n_i$ stands for the number
density of the $i$th particle species, while $\Gamma (i\rightarrow
\axino +\ldots) $ is the corresponding decay width into final states involving axinos.  Approximate
solutions for the number density of relic axinos are given in Ref.~\cite{Choi:1999xm}.

\begin{figure}[!t]
  \begin{center}
  \begin{tabular}{c}
   \includegraphics[width=0.5\textwidth]{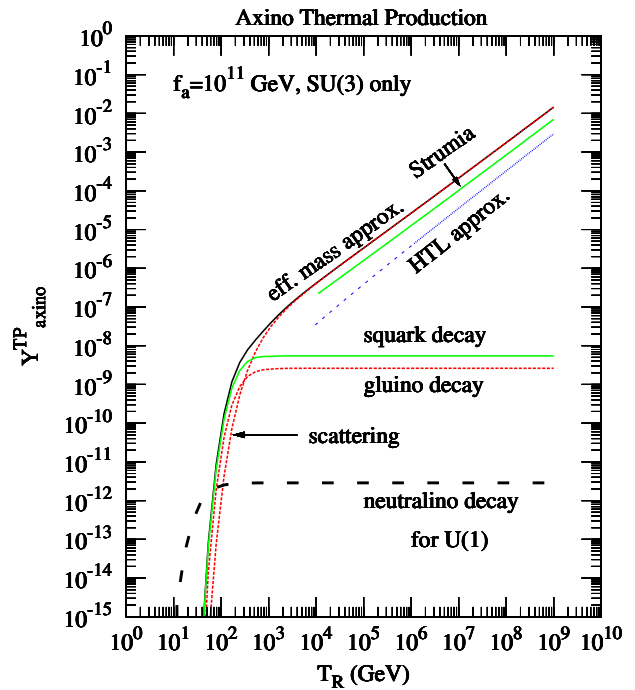}
   \end{tabular}
  \end{center}
  \caption{ Thermal yield of axino, $Y_\axino^{\rm TP}\equiv
    n_\axino/s$, versus $\treh$.  For strong interactions, the
    effective thermal mass (ETM) approximation (black) is used. We use
    the representative values of $\fa=10^{11}\gev$ and
    $m_{\squark}=\mgluino=1\tev$. For comparison, we also show the HTL
    approximation (dotted blue/dark grey) and that of Strumia
    (green/light grey).  We also denote the yield from squark (solid
    green/light grey) and gluino decay (dotted red), as well as
    out-of-equilibrium bino-like neutralino decay (dashed black) with
    $C_{aYY}=8/3$.  }
\label{fig:YTR}
\end{figure}

In Fig.~\ref{fig:YTR} (taken from Ref.~\cite{Choi:2011yf} where it was
updated from Refs.~\cite{CKKR01,CRS02}) we show the axino yield $Y$
resulting from scatterings and decays involving from strong
interactions in the KSVZ model. For different values of $\fa$, the
curves move up or down proportional to $1/\fa^2$. The contribution
from \SUW~ and \UY~ interactions are suppressed by the gauge coupling
since the cross section $\sigma \propto \alpha^3$.  (For
comparison, in Fig.~\ref{fig:YTR} we also show the yield from bino-like neutralino decay
after freezeout which is subdominant at larger $\treh$ but becomes
important at low $\treh$.)

In the case of scatterings, we compare three different prescriptions
for treating the infrared (IR) divergence that have been used in the
literature. In Ref~\cite{CKKR01} an effective thermal mass (ETM)
approximation was used to regulate the infrared divergence from
massless gluon.  A more consistent way using a hard thermal loop (HTL)
approximation was used in Ref.~\cite{Steffen04}. The technique is,
however, valid only in the regime of a small gauge coupling, $g_s\ll
1$, which corresponds to the reheating temperature $\treh \gg
10^6\gev$ where, as we shall see, axino as DM is too warm. In
Ref.~\cite{Strumia10} full resumed finite-temperature propagators for
gluons and gluinos were used which extended the validity of the
procedure down to $\treh\gsim10^4\gev$. However, the gauge invariance
in the next leading order is not maintained. We conclude that there
currently remains a factor of a few uncertainty in the thermal yield
of axinos at high $\treh$.

As noted in Ref.~\cite{Bae11}, when the temperature is higher than the
mass $M_Q$ of the PQ-charged and gauge-charged matter in the model
which induce the axino-gaugino-gauge boson interaction, the
interaction amplitude is suppressed by $M_Q^2/T^2$, in addition to the
suppression by the PQ scale $\fa$.  This is most notable in the DFSZ
model where the higgsino mass $\mu$ is around the weak scale and the
temperature is higher than this scale.

Axinos can also be efficiently produced through decays of thermal
particles via the second term in \eq{Boltzmann} when $\treh$ is
comparable to the mass of the decaying particles. At larger $\treh$
the contribution from decays becomes independent of temperature and in
any case strongly subdominant relative to that from scatterings. At
lower $\treh$ the production becomes exponentially reduced due to the
Boltzmann suppression factor for the population of the decaying
particles in the thermal plasma.

One of the decay channels is a two-body decay of a gaugino into an
axino and a gauge boson~\cite{CKKR01}.  The
gaugino-axino-sfermion-sfermion interaction in \eq{eq:Laxino}
generates three-body decays of a gaugino into an axino and two
sfermions, which is subdominant to the two-body decay.  In the KSVZ
model, an effective dimension-4 axino-quark-squark coupling is
generated at a one loop and the squark decay can produce important
amount of axinos~\cite{CRS02}. Those axino production from thermal
gluons, neutralinos and squarks are shown in Fig.~\ref{fig:YTR}.

In the DFSZ framework, the dominant
production contribution comes from scatterings involving \SUW~
interactions and also from the decays of a higgsino into an axino and
a Higgs boson due to a tree-level axino-Higgs-higgsino interaction
term~\cite{Chun:2011zd,Bae11} which is proportional to the higgsino
mass $\mu$. Axino production from higgsino decays in thermal
equilibrium is comparable to, or, for large $\mu$, can even be larger
than, that from squark decays for which a coupling exists also already
at a tree level due to the $c_2$ interaction term, which is
proportional to the mass of the quark. Generally, in the DFSZ
framework axino production from thermal decays
dominates~\cite{Chun:2011zd} over that from
scatterings~\cite{Chun:2011zd,Bae11,Bae:2011iw,CKKR01} which is
suppressed by the quark mass at higher temperature~\cite{Bae11}.
Therefore, the axino abundance is independent of the reheating
temperature, if it is high enough compared to the higgsino mass.

\vskip 0.5cm
\subsubsection{Non-thermal production}
\begin{figure}[t]
  \begin{center}
  \begin{tabular}{c}
   \includegraphics[width=0.5\textwidth]{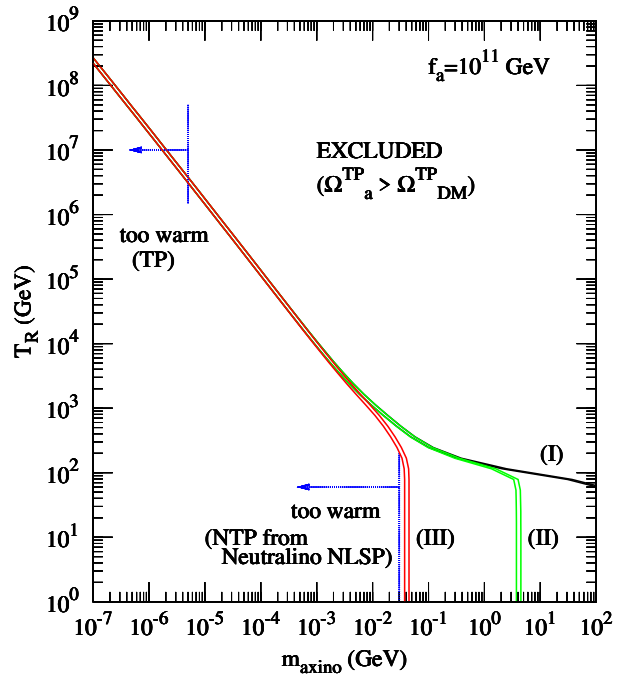}
   \end{tabular}
  \end{center}
  \caption{$\treh$ versus $\maxino$ for $\fa=10^{11}\gev$ in the KSVZ
    model. The bands inside the adjacent curves correspond to a correct
    relic density of DM axino with both TP and NTP included. The relic
    density of CDM is derived from Planck data~\cite{Ade:2013lta}
    $\Omega_{\rm CDM} h^2 = 0.1199\pm0.0027\, (68\%\, \textrm{CL})$.
    To parametrize the non-thermal production of axinos we used
    $Y_{\rm NLSP}=0$ (I), $10^{-10}$ (II), and $10^{-8}$ (III) in
    \eq{eq:NTP}. The upper right-hand area of the plot is excluded
    because of the overabundance of axinos. The regions disallowed by
    structure formation are marked with vertical blue dashed lines and
    arrows for, respectively, TP ($\maxino\lsim5\kev$) and NTP
    ($\maxino\lsim30\mev$, for a neutralino NLSP).  }
\label{fig:TR_ma}
\end{figure}

Axinos can be produced from out-of-equilibrium decays of heavier non-thermal
particles.  In this case, the axino abundance is independent of the
reheating temperature and its relic number density simply depends on the
number density and the decay modes of the decaying mother particles.

A particularly interesting case is axino production from decays of the
next-to-lightest SUSY particle (NLSP) as the lightest ordinary SUSY
particle (LOSP).  Other ordinary superpartners that are heavier than
the NLSP first cascade-decay to the NLSP which next freezes out from
thermal plasma. The NLSP then decays to the axino LSP, however, this
occurs much later on the cosmic time scale, with the lifetime around
the order of a second or less~\cite{CKR00,CKKR01}.
The
non-thermal production of axinos from the NLSP decay is given by
\dis{
\omegaantp= \frac{\maxino}{m_{\rm NLSP}} \Omega_{\rm NLSP} \simeq
2.7\times 10^{10} \bfrac{\maxino}{100\gev} Y_{\rm NLSP}.
\label{eq:NTP}
}

Cosmological axino production can also proceed from decays of other
non-thermal relics, e.g., an inflaton, moduli, a saxion,
Q-balls~\cite{Roszkowski:2006kw}, etc, however, specific
implementations are strongly model dependent and will not be
considered here.  Axino production from NLSP decay therefore provides
a conservative estimate of the relic density of axinos.

CDM relic axino production from bino decays was first considered in
Ref.~\cite{CKR00}, and next in more detail in Ref.~\cite{CKKR01} along
with thermal production and with cosmological constraints on CDM, and
also on HDM and WDM axinos. Additional processes of thermal axino
production from squark decays in thermal plasma were subsequently
considered in Ref.~\cite{CRS02}, and next applied in
Ref.~\cite{CRRS04} in an analysis of the Constrained MSSM (CMSSM) with
a neutralino and a stau as NLSP, where also analogous tau-stau-axino
couplings were obtained and applied. More recently, these couplings were
re-derived in Ref.~\cite{Wyler09} in a full two-loop calculation
including four-body hadronic decays and also for $\fa$ larger than
$10^{12}\gev$. These couplings are smaller and not important for
thermal production, but they are important for the non-thermal
production when the stau is the NLSP.  Colored NLSP was considered
in~\cite{Berger:2008ti,Covi:2009bk}, however, their contribution is
negligible due to their late freezeout.

In Fig.~\ref{fig:TR_ma} (taken from Ref.~\cite{Choi:2011yf} where it
was updated from Refs.~\cite{CKKR01,CRS02}) we consider the total
abundance of axinos (the sum of thermal and non-thermal production) to
show an upper limit on the reheating temperature for a given axino
mass from the total CDM abundance in the KSVZ model. Here we fix
$\fa=10^{11}\gev$ and the cases (I), (II) and (III) denote the
different assumed values of NLSP abundance: $Y_{\rm NLSP}=0,
10^{-10}$, and $10^{-8}$, respectively. For a small axino mass, less
than some $10\mev$, thermal production is dominant and depends on the
reheating temperature. However, for a larger mass, NTP provides the
dominant contribution. The regions above/to the right of the curves
are excluded due to the overabundance of DM.

Now we proceed to review cosmological implications of the axino
depending on its mass.

\vskip 0.5cm

\subsection{Axinos in cosmology}

If axinos are produced very late, at the time of larger than one
second after the Big Bang, from decays of an NLSP, the injection of
high energetic hadronic and electromagnetic particles can affect the
abundance of light elements produced during Big Bang Nucleosynthesis
(BBN)~\cite{Jedamzik:2004er,kkm04}.  For charged NLSP the constraints
are even stronger~\cite{Pospelov:2006sc}.  This provides an upper
bound on the abundance or the lifetime of the NLSP; especially for
large values of $\fa$ the constraint is severe as discussed
in~\cite{CKKR01,CRRS04,Wyler09}. However, as long as
$\fa\lsim10^{12}\gev$, the lifetime of bino-like NLSP in a mass range
of a few hundred GeV is less than 1 second, and for the stau is
similar, which makes axino DM free from the BBN problem.

Constraints from BBN may also be applicable when the
axino is heavy and unstable, in which case it decays into lighter MSSM
particles and SM particles~\cite{CKLS08,Bae:2013qr}.

All these scenarios depend primarily on one parameter, the axino mass,
and we discuss them in the order of increasing mass.

\vskip 0.5cm

\subsubsection{Axino as HDM}

The axino with mass in the eV and sub-eV range produced in decays of
an out-of-equilibrium sub-GeV photino was considered early on in
Ref.~\cite{Masiero84}. This case of axino can be considered as HDM.  A
primordial thermal population of axinos decoupled at $\sim
10^{11}\gev$ and was diluted away by cosmic inflation.  In $R$-parity
conserving models the photino lifetime is a function of $\fa$ and thus
the
HDM axino abundance depends on it~\cite{Masiero84}.  This
is relevant both in the standard Big Bang and the inflationary
cosmology. This is because the photino abundance is calculated from
the photino decoupling temperature which is below the reheating
temperature after inflation and hence the photino abundance is
independent of the cosmological scenarios.

A related sub-eV mass fermion useful for DM is gravitino for
$\mgravitino\lsim 1\kev$~\cite{Primack82}. Since the decoupling
temperature of gravitino is close to the Planck mass, primordial
gravitinos were diluted out in the inflationary universe. However,
axinos can decay to sub-eV gravitinos~\cite{ChunKim94}. Sub-eV
gravitinos are possible in the gauge mediated SUSY breaking
scenario. In the unstable axino case sub-eV gravitinos can
become HDM in the universe, which was called `axino-gravitino
cosmology'~\cite{KimKim95}.

\vskip 0.5cm
\subsubsection{Axino as WDM}

Rajagopal, Turner and Wilczek~\cite{RTW91} considered axinos in the
keV range. They obtained the axino mass bound $\maxino < 2\kev$ for
axinos to be WDM. However, this bound can be relaxed if the
reheating temperature $\treh$ after inflation is much lower than the
PQ symmetry breaking scale $\fa$, in which case the primordial
population of WDM axinos is diluted away and WDM axinos are subsequently
re-generated after the reheating phase.  Therefore, keV-mass
axinos can be WDM in the standard Big Bang cosmology but cannot become
WDM in the inflationary cosmology for $\treh<\fa$.  However, axinos
produced non-thermally from the decay of heavier particles can have
large free streaming length~\cite{Seto:2007ym,Choi:2012zna}.  In this
case, axinos with mass even in the $\mev$ region can be warm enough to
suppress the small scale structures that can be probed by
Lyman-$\alpha$~\cite{Boyarsky:2008xj} and
reionization~\cite{Jedamzik:2005sx} data. This is shown with blue
arrow line in Fig.~\ref{fig:TR_ma} for axinos produced from neutralino
decay.  However, this constraint is relaxed if axino population from
NTP is subdominant to the one due to DM axions.

\vskip 0.5cm
\subsubsection{Axino as CDM}
Covi, Kim and Roszkowski~\cite{CKR00} considered CDM axinos. Axinos
with mass higher than $10\kev$ for TP or $10\mev$ for NTP are
non-relativistic enough to be CDM, as marked in
Fig.~\ref{fig:TR_ma}. Even though they are relativistic at the time of
production, their velocity is red-shifted with the expansion of the
universe and they have small free streaming length at the time of
structure formation.

For CDM axinos, relatively low reheating temperatures are preferred,
as shown in Fig.~\ref{fig:TR_ma}.  Therefore, axino is a good
candidate for DM in models of thermal inflation which takes place at a late
time and naturally predicts a low reheating
temperature~\cite{Kim:2008yu,Choi:2009qd,Choi:2011rs,Jeong:2011xu,Choi:2012ye}.

Allowing for $R$-parity violation, axino DM can decay with a lifetime
much longer than the age of the universe. Photons from axino decay
can be a DM signature and can explain some astrophysical
anomalies~\cite{Hooper:2004qf,Chun:2006ss,Endo:2013si,Dey:2011zd,Hasenkamp:2011xh}.

\begin{figure}[t]
  \begin{center}
  \begin{tabular}{c}
   \includegraphics[width=0.45\textwidth]{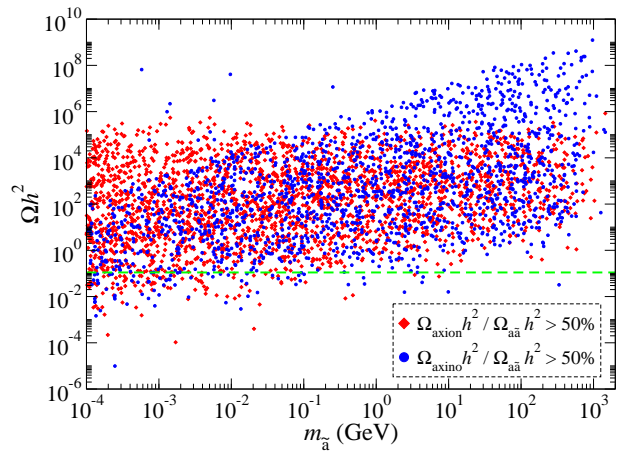}
   \end{tabular}
  \end{center}
  \caption{ The relic density of mixed axion/axino dark matter versus
    the axino mass in the scan with 19 parameters in the MSSM with
    SUSY axion multiplet.  Figure taken from Ref.~\cite{Baer:2010wm}.}
\label{fig:axionaxinoDM}
\end{figure}

Scenarios with a mixed axion-axino population of CDM have also been
considered. In Fig.~\ref{fig:axionaxinoDM} we show a numerical result
from a scan over the MSSM with 19 model parameters with a SUSY axion
model~\cite{Baer:2010wm}. In the relic density versus PQ scale plane
the dominant axion DM is shown in red and axino DM in blue.

\vskip 0.5cm

\subsubsection{Cosmology with superheavy axino}

As a digression, we note that, for super-heavy axino, it is the neutralino that
most likely is the LSP, in which case the neutralino population from
heavy axino decays could constitute CDM~\cite{CKLS08}. This case is
particularly interesting if the axino mass is greater than the
gravitino mass~\cite{Kim:2012bb} and the data from
PAMELA~\cite{PAMELA09} and recently from AMS-02~\cite{AMS13} which may
imply a TeV scale of WIMP mass, if the WIMP is
CDM. In~\cite{HuhDecay09} it was shown that the superheavy axino can
account for TeV-scale cosmic ray positrons produced as decay products
of, for example, an NMSSM singlino $\tilde N$ to $\tilde e$ plus
$e^+$, but not to antiprotons. This is possible in a string derived
flipped SU(5) grand unification model~\cite{KimKyae07}. In this case,
$\tilde e$ eventually decays to LSP neutralino plus SM particles. Of
course, the final population of the LSP is not enough to account for
the present CDM density in the decaying DM scenario but the mother
singlino density accounts for most of the CDM density, which is given
by the non-thermal production from superheavy axino decays.
Heavy axino decays to singlino and the singlino decays
  to positron and selection, and the selectron  finally decays to the
  LSP neutralino. This
$\tilde N$--WIMP decay scenario was proposed to explain the property
that the PAMELA data does not contain any large excess of antiprotons
but a significant flux of positrons~\cite{PAMELAprl09}. On the other
hand, if the decaying DM scenario is ruled out in favor of the
scattering production of positrons, the superheavy axino case can be
predominantly decaying to the LSP plus SM particles, for example in
the MSSM extended by the PQ symmetry with $S$ and $\OVER{S}$ of
Eq.~(\ref{eq:SUSYglobBr}), not introducing extra singlets $N$-type in
the NMSSM.

The decay of heavy axinos provides a non-thermal population of LSP DM
such as gravitinos or neutralinos~\cite{CKLS08}. Therefore, the
abundance of axino before decay is also constrained and gives a quite
strong limit on the reheating temperature~\cite{Cheung:2011mg}.  The
neutralino DM scenario from heavy axino decays was further studied
in~\cite{Baer_mixed,Baer:2011uz} where numerical results for the
production of mixed axion and neutralino DM were presented. The effect
of saxion production and decay was also considered and the entropy
production allowed the dilution of pre-existing particles.  The saxion
decay can also produce a sufficient population of relativistic axions which can be identified
as dark radiation in CMB observations~\cite{Choi:2012zna,Bae:2013qr}.

\begin{figure}[t]
  \begin{center}
  \begin{tabular}{c}
   \includegraphics[width=0.5\textwidth]{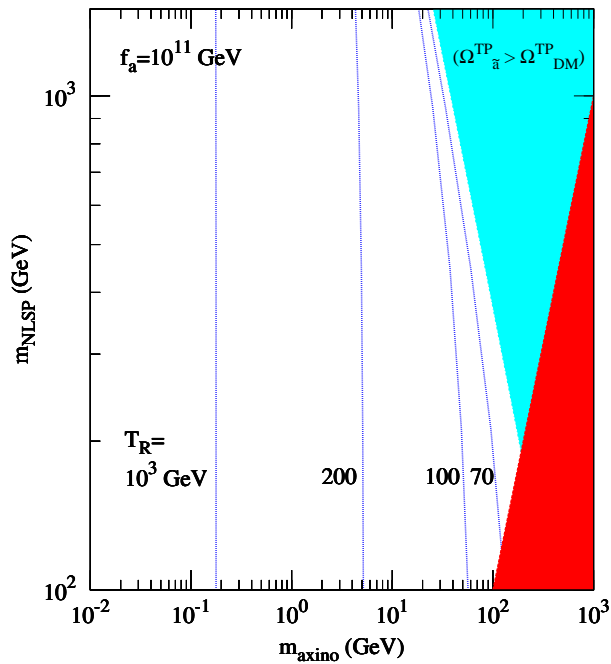}
   \end{tabular}
  \end{center}
 \caption{Contours of the reheating temperature which gives correct relic density of axino dark matter in the
    NLSP--axino mass plane. Here we have assumed $Y_{\rm
      NLSP}=10^{-12}\left({\mnlsp}/{100\gev}\right)$, typical of
    neutralino NLSP, and taken $\fa=10^{11}\gev$. The cyan wedge in the
    upper right-hand corner is excluded by the overdensity of DM,
    while in the red wedge below it the axino is not the LSP.    }
\label{fig:ms_ma_A}
\end{figure}

\section{CDM axino production at colliders}
\label{sec:AxinoColliders}
The exceedingly weak strength of axino interactions makes axino
detection in direct DM search experiments as well as at collider
experiments rather hopeless. However, at the LHC a signal of axino DM
in $R$-parity conserving scenarios may show up in the properties of the
LOSP.

With axino DM, the LOSP typically decays with the lifetime of less
than one second. Therefore, even if it is unstable on cosmological
time scales, it is practically stable inside a collider detector,
although it will decay outside it. One spectacular signature of
non-standard DM would be to detect an electrically
charged (stau) or colored (stop) massive particle as a LOSP that would
be seemingly stable in a detector.

However, this would still leave the question open whether the LSP is
the axino or for example the gravitino. Measuring the LOSP lifetime
would not be sufficient. If the LOSP is an electrically charged
particle, such as stau, then this could be possible with the analysis
of three-body decay of the charged NLSP slepton into the corresponding
lepton, the LSP and a photon. One way would be to measure the spin of the
LSP through the polarization of the final-state lepton and
photon~\cite{Buchmuller:2004rq,Buchmuller:2004tm}. Another is to
measure the branching ratio and the angular distributions of the decay
products in the three-body decays of the LOSP~\cite{Brandenburg05}.

Even if a stable massive neutral particle, such as the neutralino, is
detected at the LHC through a missing energy signature, this will
still not guarantee its DM nature conclusively since it could decay
outside of detectors. A signal indicating the same mass would
therefore be necessary in direct and indirect DM searches.  The absence of one
could indicate the existence of lighter
stable particle, such as the axino or the gravitino, playing the role
of DM.  In this way, with the axino as DM, a large region of the (C)MSSM
which would be forbidden with neutralino DM becomes
allowed~\cite{CRRS04}.

With much luck, one could in principle measure at the LHC enough
quantities to estimate, at least roughly, the relic abundance of the
LOSP and compare it with the cosmological value. If the two were
radically different, again this would indicate a non-standard choice
for DM. A model-independent study of axino DM from collider
information was performed in Ref.~\cite{ChoiKY07}.  Based on the
thermal and non-thermal production of axinos and the cold DM density,
\dis{
\Omega_{\rm \axino}^{\rm TP}  h^2 &(\treh, \maxino,m_{\tilde{g}},m_{\rm LOSP} ,\ldots)
 + \frac{\maxino}{m_{\rm LOSP} } \Omega_{\rm LOSP}h^2\\
&\quad \quad \quad = \Omega_{\rm CDM}h^2\simeq 0.1.
}
and from the information inferred on the relic density of LOSP at
colliders, one could attempt to determine a relation between the reheating
temperature and the mass of axino. This is shown in
Fig.~\ref{fig:TR_ma} (taken from Ref.~\cite{ChoiKY07}).

\section{Conclusions}
\label{sec:conclusions}
Axino is an intriguing candidate for DM in SUSY with the axion
solution of the strong $CP$ problem.  The mass and interactions are
highly dependent on the axion model and its coupling to SM particles
are suppressed by the PQ scale $\fa$. However, SUSY breaking alters
the connection so that the axino becomes massive and can also have
mixing with the goldstino~\cite{Kim:2012bb}.  Axinos can be produced
in the thermal plasma in the early Universe in scatterings and decays
of heavier particles and in out-of-equilibrium decays of heavier
particles. Axino relic density can be obtained to coincide with the
correct density of DM. Axinos can be HDM, WDM or CDM depending on
their mass range and production mechanism. There could also be two
populations of axinos, for example one warm and one cold, from different
production modes. In the case of heavy axinos, neutralinos can be
produced from decays of axinos and extend the available parameter
space and help to explain the positron anomaly in PAMELA and
AMS-02. The identification of LOSP and its interactions at colliders
might provide a relationship between the reheating temperature and the
axino mass, and also glimpses into the earliest moments of the Universe after
inflation if the DM is made up of thermally produced axinos.

\section*{Acknowledgments}
K.-Y.C. was supported by the Basic Science Research Program through
the National Research Foundation of Korea (NRF) funded by the Ministry
of Education, Science and Technology Grant No. 2011-0011083.
K.-Y.C. acknowledges the Max Planck Society (MPG), the Korea Ministry
of Education, Science and Technology (MEST), Gyeongsangbuk-Do and
Pohang City for the support of the Independent Junior Research Group
at the Asia Pacific Center for Theoretical Physics (APCTP). J.E.K. is
supported in part by the National Research Foundation (NRF) grant
funded by the Korean Government (MEST) (No. 2005-0093841). L.R. is
supported by the Welcome Programme of the Foundation for Polish
Science.

%


\def\prp#1#2#3{Phys.\ Rep.\ {\bf #1} (#3) #2}
\def\rmp#1#2#3{Rev. Mod. Phys.\ {\bf #1} (#3) #2}
\def\anrnp#1#2#3{Annu. Rev. Nucl. Part. Sci.\ {\bf #1} (#3) #2}
\def\npb#1#2#3{Nucl.\ Phys.\ {\bf B#1} (#3) #2}
\def\plb#1#2#3{Phys.\ Lett.\ {\bf B#1} (#3) #2}
\def\prd#1#2#3{Phys.\ Rev.\ {\bf D#1}, #2 (#3)}
\def\prl#1#2#3{Phys.\ Rev.\ Lett.\ {\bf #1} (#3) #2}
\def\jhep#1#2#3{JHEP\ {\bf #1} (#3) #2}
\def\jcap#1#2#3{JCAP\ {\bf #1} (#3) #2}
\def\zp#1#2#3{Z.\ Phys.\ {\bf #1} (#3) #2}
\def\epjc#1#2#3{Euro. Phys. J.\ {\bf #1} (#3) #2}
\def\ijmp#1#2#3{Int.\ J.\ Mod.\ Phys.\ {\bf #1} (#3) #2}
\def\mpl#1#2#3{Mod.\ Phys.\ Lett.\ {\bf #1} (#3) #2}
\def\apj#1#2#3{Astrophys.\ J.\ {\bf #1} (#3) #2}
\def\nat#1#2#3{Nature\ {\bf #1} (#3) #2}
\def\sjnp#1#2#3{Sov.\ J.\ Nucl.\ Phys.\ {\bf #1} (#3) #2}
\def\apj#1#2#3{Astrophys.\ J.\ {\bf #1} (#3) #2}
\def\ijmp#1#2#3{Int.\ J.\ Mod.\ Phys.\ {\bf #1} (#3) #2}
\def\apph#1#2#3{Astropart.\ Phys.\ {\bf B#1}, #2 (#3)}
\def\mnras#1#2#3{Mon.\ Not.\ R.\ Astron.\ Soc.\ {\bf #1} (#3) #2}
\def\nat#1#2#3{Nature (London)\ {\bf #1} (#3) #2}
\def\apjs#1#2#3{Astrophys.\ J.\ Supp.\ {\bf #1} (#3) #2}
\def\aipcp#1#2#3{AIP Conf.\ Proc.\ {\bf #1} (#3) #2}





\begin{thebibliography}{99}


\bibitem{CKR00}  L.~Covi, J.~E.~Kim and L.~Roszkowski,
  \prl{82}{4180}{1999}  [hep-ph/9905212].

\bibitem{CKKR01}
  L.~Covi, H.~B.~Kim, J.~E.~Kim and L.~Roszkowski,
  \jhep{0105}{033}{2001}  [hep-ph/0101009].

\bibitem{CRS02}
  L.~Covi, L.~Roszkowski and M.~Small,
  \jhep{0207}{023}{2002}  [hep-ph/0206119].

\bibitem{CRRS04}
  L.~Covi, L.~Roszkowski, R.~Ruiz de Austri and M.~Small,
  \jhep{0406}{003}{2004} [hep-ph/0402240].

\bibitem{Steffen04}
  A.~Brandenburg and F.~D.~Steffen,
  \jcap{0408}{008}{2004}  [hep-ph/0405158].

\bibitem{Strumia10}
  A.~Strumia,
  \jhep{1006}{036}{2010}  [arXiv:1003.5847 [hep-ph]].

\bibitem{AxinoRevs} For a recent review, see,  L. Covi and J. E. Kim, New J. of Phys. {\bf 11} (2009) 105003 
    [arXiv: 0902.0769[astro-ph/CO]].

\bibitem{Wyler09}
A. Freitas, F. D. Steffen, N. Tajuddin, and D. Wyler, \plb{679}{270}{2009} [arXiv:0904.3218[hep-ph]];
A. Freitas, F. D. Steffen, N. Tajuddin, and D. Wyler, \plb{682}{193}{2009} [arXiv:0909.3293[hep-ph]].
  A.~Freitas, F.~D.~Steffen, N.~Tajuddin and D.~Wyler,  \jhep{1106}{036}{2011}
  [arXiv:1105.1113 [hep-ph]].

\bibitem{flaxino}
  E.~J.~Chun, H.~B.~Kim, K.~Kohri, D.~H.~Lyth,
  \jhep{0803}{061}{2008} [arXiv:0801.4108 [hep-ph]];
  S.~Kim, W.~I.~Park and E.~D.~Stewart,
  \jhep{0901}{015}{2009}  [arXiv:0807.3607 [hep-ph]].


\bibitem{Baer}
 H.~Baer and A.~D.~Box,
  Eur.\ Phys.\ J.\ C {\bf 68} (2010) 523
  [arXiv:0910.0333 [hep-ph]];
  H.~Baer, R.~Dermisek, S.~Rajagopalan, H.~Summy,
  \jcap{1007}{014}{2010}
  [arXiv:1004.3297 [hep-ph]];
  H.~Baer, S.~Kraml, A.~Lessa, S.~Sekmen,
  \jcap{1011}{040}{2010}
  [arXiv:1009.2959 [hep-ph]];
  H.~Baer, S.~Kraml, A.~Lessa, S.~Sekmen,
  \jcap{1104}{039}{2011}
  [arXiv:1012.3769 [hep-ph]];

\bibitem{Baer:2010wm}
  H.~Baer, A.~D.~Box, H.~Summy,
  \jhep{1010}{023}{2010}
  [arXiv:1005.2215 [hep-ph]];

  \bibitem{Baer_mixed}
   H.~Baer, A.~Lessa, S.~Rajagopalan and W.~Sreethawong,
  \jcap{1106}{031}{2011}
  [arXiv:1103.5413 [hep-ph]];
  H.~Baer and A.~Lessa,
  \jhep{1106}{027}{2011}
  [arXiv:1104.4807 [hep-ph]];

\bibitem{Baer:2011uz}
  H.~Baer, A.~Lessa and W.~Sreethawong,
  \jcap{1201}{036}{2012}
  [arXiv:1110.2491 [hep-ph]].

\bibitem{Kang:2008jq}
  J.~UKang and G.~Panotopoulos,
 \jhep{0805}{036}{2008}
  [arXiv:0805.0535 [hep-ph]].


\bibitem{NillesRaby82} H. P. Nilles and S. Raby, \npb{198}{102}{1982}.

\bibitem{Tamv82} K. Tamvakis and D. Wyler, \plb{112}{451}{1982}.

\bibitem{Frere83} J. M. Frere and J. M. Gerard, Lett. Nuovo Cim. {\bf 37}, (1983) 135.

\bibitem{KimRMP10} For a recent review, see, J. E. Kim and G. Carosi, \rmp{82}{557}{2010} [arXiv: 0807.3125[hep-ph]].

\bibitem{Nilles84} For a review, see, H. P. Nilles, \prp{110}{1}{1984}.

\bibitem{ChunKN92} E. J. Chun, J. E. Kim and H. P. Nilles, \plb{287}{123}{1992}.

\bibitem{Chun:1995hc}
  E.~J.~Chun and A.~Lukas,
  Phys.\ Lett. {\bf B 357} (1995) 43
  [hep-ph/9503233].


\bibitem{Kim:2012bb}
  J.~E.~Kim and M.~-S.~Seo,
  Nucl.\ Phys. {\bf B864} (2012) 296
  [arXiv:1204.5495 [hep-ph]].

\bibitem{Masiero84} J. E. Kim, A. Masiero and D. V. Nanopoulos, \plb{139}{346}{1984}.

\bibitem{RTW91}
K.~Rajagopal, M.S.~Turner and F.~Wilczek,
\npb{358}{447}{1991}.



\bibitem{BaeHuhKim09} K.-J. Bae, J.-H. Huh and J. E. Kim, \jcap{0809}{005}{2009} [arXiv:0806.0497 [hep-ph]].


\bibitem{KSVZ79} J. E. Kim, \prl{43}{103}{1979}; M. A. Shifman, V. I. Vainstein, V. I. Zakharov, \npb{166}{4933}{1980}.

\bibitem{DFSZ81} M. Dine, W. Fischler and M. Srednicki, \plb{104}{199}{1981};  A. P. Zhitnitskii, Sov. J. Nucl. Phys. {\bf 31} (1980) 260.

\bibitem{Kim98}
  J.~E.~Kim,
  Phys.\ Rev. {\bf D58} (1998) 055006
  [arXiv:hep-ph/9802220].

\bibitem{IWKim06} K.-S. Choi, I.-W. Kim and J. E. Kim, \jhep{0703}{116}{2007} [arXiv:hep-ph/0612107].


\bibitem{Kim83} J. E. Kim, \plb{136}{78}{1984}.


\bibitem{Kim13} J. E. Kim, \prl{111}{}{2013} [arXiv:1303.1822 [hep-ph]].


\bibitem{Barr92}
S. M. Barr and D. Seckel, 
     \prd{46}{1992}{539};
M. Kamionkowski and J. March-Russell,
          \plb{282}{137}{1992} [hep-th/9202003];
R. Holman, S. D. H. Hsu, T. W. Kephart, E. W. Kolb, R. Watkins, and L. M. Widrow,
    \plb{282}{132}{1992} [hep-ph/9203206];
S. Ghigna, M. Lusignoli and M. Roncadelli, 
   \plb{283}{278}{1992};
B. A. Dobrescu, 
   \prd{55}{1997}{5826} [hep-ph/9609221].



\bibitem{Sikivie:1982qv}
  P.~Sikivie,
  Phys.\ Rev.\ Lett.\  {\bf 48} (1982) 1156.

\bibitem{Choi:2011yf}
  K.~-Y.~Choi, L.~Covi, J.~E.~Kim and L.~Roszkowski,
  \jhep{1204}{106}{2012}
  [arXiv:1108.2282 [hep-ph]].

\bibitem{Bae11}
  K.~J.~Bae, K.~Choi and S.~H.~Im,
  \jhep{1108}{065}{2011}
  [arXiv:1106.2452 [hep-ph]].

\bibitem{Nieves:1985fq}
  J.~F.~Nieves,
  Phys.\ Rev. {\bf D33} (1986) 1762.

\bibitem{Goto:1991gq}
  T.~Goto and M.~Yamaguchi,
  Phys.\ Lett. {\bf B276} (1992) 103.

\bibitem{Moxhay:1984am}
  P.~Moxhay and K.~Yamamoto,
  Phys.\ Lett. {\bf B151} (1985) 363.

\bibitem{Kim91} J. E. Kim, \prl{67}{3465}{1991}.



\bibitem{KimHB96} S. Chang and H. B. Kim, \prl{77}{591}{1996} [arXiv:hep-ph/9604222].

\bibitem{Kawasaki:2007mk}
  M.~Kawasaki, K.~Nakayama and M.~Senami,
 \jcap{0803}{009}{2008}
  [arXiv:0711.3083 [hep-ph]].

\bibitem{Kawasaki:2011ym}
  M.~Kawasaki, N.~Kitajima and K.~Nakayama,
  Phys.\ Rev. {\bf D83} (2011) 123521
  [arXiv:1104.1262 [hep-ph]].

\bibitem{Choi:2012zna}
  K.~Choi, K.~-Y.~Choi and C.~S.~,
  Phys.\ Rev. {\bf D86} (2012) 083529
  [arXiv:1208.2496 [hep-ph]].

\bibitem{Hasenkamp:2011em}
  J.~Hasenkamp,
  Phys.\ Lett. {\bf B707} (2012) 121
  [arXiv:1107.4319 [hep-ph]].

\bibitem{Graf:2012hb}
  P.~Graf and F.~D.~Steffen,
  \jcap{1302}{018}{2013}
  [arXiv:1208.2951 [hep-ph]].

\bibitem{Bae:2013qr}
  K.~J.~Bae, H.~Baer and A.~Lessa,
  \jcap{1304}{041}{2013}
  [arXiv:1301.7428 [hep-ph]].

\bibitem{Graf:2013xpe}
  P.~Graf and F.~D.~Steffen,
  arXiv:1302.2143 [hep-ph].



\bibitem{CMSSM-dm} L.~Roszkowski,
{\it Pramana} {\bf 62} (2004) 389 [arXiv:hep-ph/0404052];

\bibitem{Boyarsky:2005us}
  A.~Boyarsky, A.~Neronov, O.~Ruchayskiy and M.~Shaposhnikov,
  Mon.\ Not.\ Roy.\ Astron.\ Soc.\  {\bf 370} (2006) 213
  [astro-ph/0512509].

\bibitem{Boyarsky:2009ix}
  A.~Boyarsky, O.~Ruchayskiy and M.~Shaposhnikov,
  Ann.\ Rev.\ Nucl.\ Part.\ Sci.\  {\bf 59} (2009) 191
  [arXiv:0901.0011 [hep-ph]].


\bibitem{Abazajian:2012ys}
  K.~N.~Abazajian, M.~A.~Acero, S.~K.~Agarwalla, A.~A.~Aguilar-Arevalo, C.~H.~Albright, S.~Antusch, C.~A.~Arguelles and A.~B.~Balantekin {\it et al.},
  arXiv:1204.5379 [hep-ph].



\bibitem{Ellis:1984eq}
  J.~R.~Ellis, J.~E.~Kim, D.~V.~Nanopoulos,
  Phys.\ Lett.\  {\bf B145 } (1984)  181.
E. Holtmann, M. Kawasaki, K. Kohri, and T. Moroi, \prd{60}{023506}{1999} [hep-ph/9805405].

\bibitem{Kawasaki:1994af}
  M.~Kawasaki and T.~Moroi,
  Prog.\ Theor.\ Phys.\  {\bf 93} (1995) 879
  [hep-ph/9403364, hep-ph/9403061].

\bibitem{Gomez:2008js}
  M.~E.~Gomez, S.~Lola, C.~Pallis and J.~Rodriguez-Quintero,
  \jcap{0901}{027}{2009}
  [arXiv:0809.1859 [hep-ph]].

\bibitem{Choi:1999xm}
  K.~Choi, K.~Hwang, H.~B.~Kim, T.~Lee,
  Phys.\ Lett.\  {\bf B467 } (1999)  211-217.
  [hep-ph/9902291].

\bibitem{Chun:2011zd}
  E.~J.~Chun,
  Phys.\ Rev. {\bf D84} (2011) 043509
  [arXiv:1104.2219 [hep-ph]].

\bibitem{Bae:2011iw}
  K.~J.~Bae, E.~J.~Chun and S.~H.~Im,
  \jcap{1203}{013}{2012}
  [arXiv:1111.5962 [hep-ph]].

\bibitem{Roszkowski:2006kw}
  L.~Roszkowski, O.~Seto,
  Phys.\ Rev.\ Lett.\  {\bf 98 } (2007)  161304.
  [hep-ph/0608013].


\bibitem{Ade:2013lta}
  P.~A.~R.~Ade {\it et al.}  [ Planck Collaboration],
  arXiv:1303.5076 [astro-ph.CO].



\bibitem{Berger:2008ti}
  C.~F.~Berger, L.~Covi, S.~Kraml and F.~Palorini,
  \jcap{0810}{005}{2008}
  [arXiv:0807.0211 [hep-ph]].

\bibitem{Covi:2009bk}
  L.~Covi, J.~Hasenkamp, S.~Pokorski and J.~Roberts,
  \jhep{0911}{003}{2009}
  [arXiv:0908.3399 [hep-ph]].



\bibitem{Jedamzik:2004er}
  K.~Jedamzik,
  Phys.\ Rev. {\bf D70} (2004) 063524
  [arXiv:astro-ph/0402344].

\bibitem{kkm04}
M.~Kawasaki, K.~Kohri and T.~Moroi,
\plb{625}{2005}{7} [astro-ph/0402490];
\prd{71}{2005}{083502} [astro-ph/0408426].


\bibitem{Pospelov:2006sc}
  M.~Pospelov,
  \prl{98}{2007}{231301} [hep-ph/0605215].



\bibitem{CKLS08}  K.-Y. Choi, J. E. Kim, H. M. Lee, and O. Seto, \prd{77}{123501}{2008} 
    [arXiv: 0801.0491[hep-ph]].

\bibitem{Primack82} H. Pagels and J. R. Primack, \prl{48}{223}{1982}.


\bibitem{ChunKim94} E.  J.  Chun, H. B. Kim and J. E. Kim,  \prl{72}{1956}{1994} [arXiv:hep-ph/9305208].

\bibitem{KimKim95}  H. B. Kim and J. E. Kim, \npb{433}{421}{1995} [arXiv:hep-ph/9405385].


\bibitem{Seto:2007ym}
  O.~Seto and M.~Yamaguchi,
  Phys.\ Rev. {\bf D75} (2007) 123506
  [arXiv:0704.0510 [hep-ph]].



\bibitem{Boyarsky:2008xj}
  A.~Boyarsky, J.~Lesgourgues, O.~Ruchayskiy and M.~Viel,
  \jcap{0905}{012}{2009}
  [arXiv:0812.0010 [astro-ph]].


\bibitem{Jedamzik:2005sx}
  K.~Jedamzik, M.~Lemoine and G.~Moultaka,
  \jcap{0607}{010}{2006}
  [arXiv:astro-ph/0508141].
\bibitem{Kim:2008yu}
  S.~Kim, W.~-I.~Park and E.~D.~Stewart,
  \jhep{0901}{015}{2009}
  [arXiv:0807.3607 [hep-ph]].

\bibitem{Choi:2009qd}
  K.~Choi, K.~S.~Jeong, W.~-I.~Park and C.~S.~Shin,
  \jcap{0911}{018}{2009}
  [arXiv:0908.2154 [hep-ph]].


\bibitem{Choi:2011rs}
  K.~Choi, E.~J.~Chun, H.~D.~Kim, W.~I.~Park and C.~S.~Shin,
  Phys.\ Rev. {\bf D83} (2011) 123503
  [arXiv:1102.2900 [hep-ph]].

\bibitem{Jeong:2011xu}
  K.~S.~Jeong and M.~Yamaguchi,
  \jhep{1107}{124}{2011}
  [arXiv:1102.3301 [hep-ph]].

\bibitem{Choi:2012ye}
  K.~Choi, W.~-I.~Park and C.~S.~Shin,
  \jcap{1303}{011}{2013}
  [arXiv:1211.3755 [hep-ph]].

\bibitem{Hooper:2004qf}
  D.~Hooper and L.~-T.~Wang,
  \prd{70}{063506}{2004}
  [hep-ph/0402220].
\bibitem{Chun:2006ss}
  E.~J.~Chun and H.~B.~Kim,
  \jhep{0610}{082}{2006}
  [hep-ph/0607076].
\bibitem{Endo:2013si}
  M.~Endo, K.~Hamaguchi, S.~P.~Liew, K.~Mukaida and K.~Nakayama,
  Phys.\ Lett. {\bf B721} (2013) pp. 111
  [arXiv:1301.7536 [hep-ph]].
\bibitem{Dey:2011zd}
  P.~Dey, B.~Mukhopadhyaya, S.~Roy and S.~K.~Vempati,
  \jcap{1205}{042}{2012}
  [arXiv:1108.1368 [hep-ph]].
\bibitem{Hasenkamp:2011xh}
  J.~Hasenkamp and J.~Kersten,
  Phys.\ Lett.  {\bf B701} (2011) 660
  [arXiv:1103.6193 [hep-ph]].


\bibitem{PAMELA09} O. Adriani \etal (PAMELA Collaboration), \nat{458}{607}{2009} [arXiv:0810.4995[astro-ph]].

\bibitem{AMS13} M. Aguilar \etal (AMS Collaboration), \prl{110}{141102}{2013}.

\bibitem{HuhDecay09} J.-H. Huh and J. E. Kim, \prd{80}{075012}{2009} [arXiv: 0908.0152[hep-ph]].

\bibitem{KimKyae07} J. E. Kim and B. Kyae, \npb{770}{47}{2007} [arXiv:hep-th/0608086].


\bibitem{PAMELAprl09} O. Adriani \etal (PAMELA Collaboration), \prl{102}{051101}{2009} [arXiv:0810.4994[astro-ph]].

\bibitem{Cheung:2011mg}
  C.~Cheung, G.~Elor and L.~J.~Hall,
  Phys.\ Rev. {\bf D85} (2012) 015008
  [arXiv:1104.0692 [hep-ph]].


\bibitem{Buchmuller:2004rq}
  W.~Buchmuller, K.~Hamaguchi, M.~Ratz and T.~Yanagida,
  Phys.\ Lett. {\bf B588} (2004) 90
  [hep-ph/0402179].
\bibitem{Buchmuller:2004tm}
  W.~Buchmuller, K.~Hamaguchi, M.~Ratz and T.~Yanagida,
  hep-ph/0403203.

\bibitem{Brandenburg05}
  A.~Brandenburg, L.~Covi, K.~Hamaguchi, L.~Roszkowski and F.~D.~Steffen,
  Phys.\ Lett. {\bf B617} (2005) 99
  [arXiv:hep-ph/0501287].



\bibitem{ChoiKY07}
  K.~Y.~Choi, L.~Roszkowski and R.~Ruiz de Austri,
  \jhep{0804}{016}{2008}
  [arXiv:0710.3349 [hep-ph]].






\end{thebibliography}
\end{document}